\documentclass[conference]{IEEEtran}
\IEEEoverridecommandlockouts
\usepackage{cite}
\usepackage{amsmath,amssymb,amsfonts}
\usepackage{algorithmic}
\usepackage{adjustbox}
\usepackage{booktabs}
\usepackage{graphicx}
\usepackage[para,online,flushleft]{threeparttable}
\usepackage{textcomp}
\usepackage{multirow}
\usepackage{float}
\usepackage{caption}
\usepackage[para,online,flushleft]{threeparttable}
\usepackage[dvipsnames]{xcolor}
\def\BibTeX{{\rm B\kern-.05em{\sc i\kern-.025em b}\kern-.08em
    T\kern-.1667em\lower.7ex\hbox{E}\kern-.125emX}}

\begin{document}

\bstctlcite{BSTcontrol}

\title{Hummingbird: A Smaller and Faster Large Language Model Accelerator on Embedded FPGA}

\author{\normalsize
Jindong Li\textsuperscript{\scriptsize *1,2,3,6} \enspace
Tenglong Li\textsuperscript{\scriptsize *1,2,3,6} \enspace
Ruiqi Chen\textsuperscript{\scriptsize 4} \enspace
Guobin Shen\textsuperscript{\scriptsize 1,2,3,5} \enspace
Dongcheng Zhao\textsuperscript{\scriptsize 2,3} \enspace
Qian Zhang\textsuperscript{\scriptsize 1,2,3,6} \enspace
Yi Zeng\textsuperscript{\scriptsize 1,2,3,5,6,7}\\
$^1$Brain-inspired Cognitive Intelligence Lab, Institute of Automation, Chinese Academy of Sciences\\ 
$^2$ Center for Long-term Artificial Intelligence,
$^3$ Beijing Institute of AI Safety and Governance,
$^4$ Vrije Universiteit Brussel\\
$^5$ School of Future Technology,
$^6$ School of Artificial Intelligence, University of Chinese Academy of Sciences  \\ 
$^7$ Key Laboratory of Brain Cognition and Brain-inspired Intelligence Technology, Chinese Academy of Sciences\\ 
{\{lijindong2022, litenglong2023, shenguobin2021,}{zhaodongcheng2016, q.zhang, yi.zeng\}@ia.ac.cn, ruiqi.chen@vub.be}
\thanks{*Equal Contribution. Corresponding author: Qian Zhang and Yi Zeng.}
}

\maketitle

\begin{abstract}

Deploying large language models (LLMs) on embedded devices remains a significant research challenge due to the high computational and memory demands of LLMs and the limited hardware resources available in such environments. While embedded FPGAs have demonstrated performance and energy efficiency in traditional deep neural networks, their potential for LLM inference remains largely unexplored.
Recent efforts to deploy LLMs on FPGAs have primarily relied on large, expensive cloud-grade hardware and have only shown promising results on relatively small LLMs, limiting their real-world applicability.
In this work, we present \textit{Hummingbird}, a novel FPGA accelerator designed specifically for LLM inference on embedded FPGAs.
Hummingbird is \textit{smaller}—targeting embedded FPGAs such as the KV260 and ZCU104 with 67\% LUT, 39\% DSP, and 42\% power savings over existing research.
Hummingbird is \textit{stronger}—targeting LLaMA3-8B and supporting longer contexts, overcoming the typical 4GB memory constraint of embedded FPGAs through offloading strategies.
Finally, Hummingbird is \textit{faster}—achieving 4.8 tokens/s and 8.6 tokens/s for LLaMA3-8B on the KV260 and ZCU104 respectively, with 93-94\% model bandwidth utilization, outperforming the prior 4.9 token/s for LLaMA2-7B with 84\% bandwidth utilization baseline.
We further demonstrate the viability of industrial applications by deploying Hummingbird on a cost-optimized Spartan UltraScale FPGA, paving the way for affordable LLM solutions at the edge.


\end{abstract}


\section{Introduction}

Large Language Models (LLMs) have unlocked powerful capabilities in natural language understanding and generation, powering applications such as chatbots, code assistants and machine translation. While cloud-based infrastructure provides a scalable solution for many of these workloads, certain use cases demand real-time responsiveness that the cloud cannot guarantee. In applications such as privacy-aware chat interfaces and embodied AI systems, it is crucial to support LLM for individual user queries directly on edge devices.

\begin{figure}
  \centering
  \includegraphics[width=1.0\linewidth]{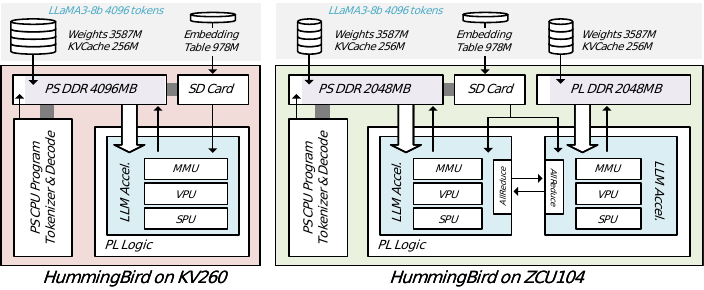}
  \caption{Hummingbird system on two embedded FPGA platforms.}
  \label{fig:firstpic}
  \end{figure}

Field Programmable Gate Arrays (FPGAs) have long been recognized as an efficient edge solution for accelerating deep learning workloads\cite{guo2017angel}\cite{li2024firefly}\cite{li2024fireflys}. However, the explosive growth in the scale of LLM presents an unprecedented challenge, deploying LLM onto embedded FPGA has become more daunting than ever before.
Existing research has primarily focused on deploying relatively small LLM on high-end, cloud FPGAs.
One of the earliest efforts, DFX \cite{hong2022dfx}, utilized four Alveo U280s—each equipped with High Bandwidth Memory (HBM)—to accelerate the GPT-2 1.5B model. Chen et al. \cite{chen2024understanding} further demonstrated the advantages of FPGAs during the decoding phase by deploying GPT-2 1.5B on a single Alveo U280.
Among the more recent works, FlightLLM \cite{zeng2024flightllm} successfully deployed LLaMA2-7B on a single Alveo U280, representing the largest LLM deployed on a single FPGA to date. EdgeLLM \cite{huang2025edgellm} further enhanced performance by deploying ChatGLM-6B on the VCU128, another HBM device.
LoopLynx\cite{zheng2025looplynx} deployed GPT2-345M on Alveo U50.
Meanwhile, ChatOPU \cite{chatopu} implemented OPT-1.3B on a DDR-based Alveo U200, and LightMamba \cite{wei2025lightmamba} deployed Mamba2-2.7B on the VCK190.
Notably, all these efforts have concentrated on cloud FPGAs with model sizes capped at 7B, and all these works achieve limited bandwidth utilization——the most critical resource for memory-bound LLM inference.

A notable exception is the work by Li et al. \cite{li2025pushing}, which represents the first and, to date, only attempt to deploy LLaMA2-7B\cite{touvron2023llama2openfoundation} on an embedded KV260 platform. It pushes up to the limit of embedded FPGA inference, achieving a decoding speed of 4.9 token/s by utilizing 84\% of the available bandwidth. Furthermore, 93\% of the memory capacity was utilized for model weights and $kv$ cache, enabling inference of sequences up to 1K tokens—insufficient for most applications.

\begin{figure*}
  \centering
  \includegraphics[width=1.0\linewidth]{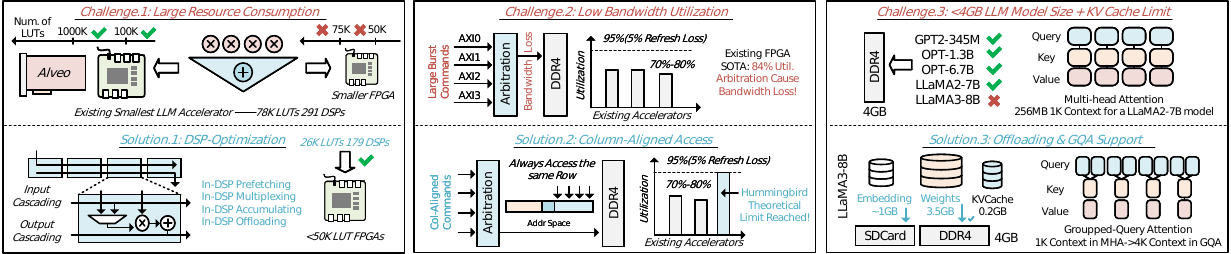}
  \caption{Existing inefficiencies in LLM accelerator on embedded FPGA and proposed improvements.}
  \label{fig:method}
  \end{figure*}

Despite these advancements, several limitations remain. This work focuses on the following three research questions:
(1) How can the resource consumption of LLM accelerators be further reduced to enable deployment on \textbf{\textit{smaller}}, embedded FPGAs with lower cost?
(2) How can we achieve \textbf{\textit{faster}} decoding throughput and higher memory bandwidth utilization, while maintaining an expandable architecture that can leverage additional memory resources?
(3) How can we deploy models that are larger and \textbf{\textit{stronger}} than LLaMA2-7B, while supporting longer token sequences, without relying on further compression techniques that degrade accuracy?

In this work, we propose \textbf{\textit{Hummingbird}}, a smaller and faster LLM accelerator designed to support stronger LLM and longer contexts—specifically, the more advanced LLaMA3-8B\cite{grattafiori2024llama3herdmodels}, on embedded FPGAs.
The name Hummingbird reflects the architecture's key features: small and fast, mirroring the bird's distinctive traits. Additionally, the term humming metaphorically points to chatbot applications powered by LLM.
Building on this philosophy, our contributions are as follows:

1) We design a DSP-efficient compute engine that supports flexible GEMV operations as the core of the architecture. This optimization yields 67\% LUT, 39\% DSP, and 42\% power savings over existing LLM accelerators on embedded FPGAs, enabling deployment on \textbf{\textit{smaller}} FPGA devices.

2) We identify the primary bandwidth utilization bottleneck as the arbitration of AXI ports in the Zynq memory controller. To address this, we propose a column-aligned memory access strategy, improving model bandwidth utilization from the SOTA 84\% to 93-94\%, achieving 4.8 tokens/s and 8.6 tokens/s for LLaMA3-8B on the KV260 and ZCU104 respectively, \textbf{\textit{faster}} than existing 4.9 token/s for LLaMA2-7B baseline.

3) To accommodate the \textbf{\textit{stronger}} LLaMA3-8B model, we offload the storage-intensive embedding table to external flash memory. Additionally, we design an efficient dataflow that supports Grouped-Query Attention (GQA)\cite{ainslie2023gqa}, which can alleviate the storage pressure of the $kv$ cache and extend the supported context length to 4096 tokens.

The Hummingbird system is illustrated in Fig.\ref{fig:firstpic}, and the proposed optimization techniques are summarized in Fig.\ref{fig:method}.

\section{Preliminaries}


\subsection{Model Size Breakdown of LLM}

\begin{figure}[h]
  \centering
  \includegraphics[width=1.0\linewidth]{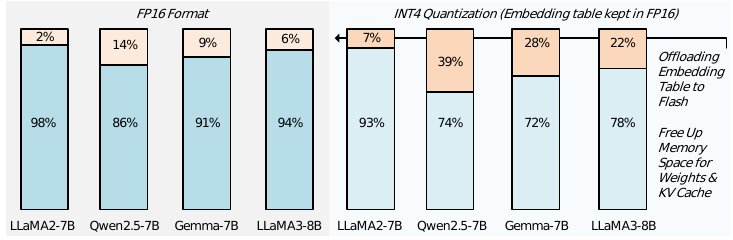}
  \caption{FP16 and INT4 quantized model size breakdowns show that the embedding table occupies a significantly larger proportion in quantized models, necessitating offloading strategy to free up memory space.}
  \label{fig:breakdown}
  \end{figure}

Limited memory capacity is the primary challenge in deploying larger LLMs on embedded FPGAs. Prior research \cite{li2025pushing} has demonstrated 7B model represents the upper bound of what embedded devices with 4GB of memory can accommodate under 4-bit quantization. To support further analysis, we provide a detailed breakdown of model size across several 7B-scale LLMs\cite{touvron2023llama2openfoundation}\cite{grattafiori2024llama3herdmodels}\cite{yang2024qwen2}\cite{team2024gemma}.
We focus on analyzing the size of the embedding table.
When the model is stored in FP16, the embedding table typically accounts for $<$10\% of the total model size. However, under quantization, the embedding table is often retained in FP16 to preserve model accuracy. As a result, its relative contribution to the overall model size increases, as shown in Fig.\ref{fig:breakdown}. Furthermore, recent LLM architectures including LLaMA3\cite{grattafiori2024llama3herdmodels}, Qwen2.5\cite{yang2024qwen2} and Gemma\cite{team2024gemma} tend to adopt larger vocabulary sizes to enhance performance, further inflating the size of the embedding table.

During a single decoding pass in LLM inference, all model weights must be accessed once, whereas only a single vector from the embedding table is required.
Therefore, we argue that the embedding table can be offloaded to a slower external memory without significantly impacting performance. This method frees up valuable on-chip memory resources, enabling more space for model weights and $kv$ cache. As a result, models exceeding the 4GB limit—such as LLaMA3-8B—can be deployed on 4GB devices with embedding table offloaded.
For models like LLaMA2-7B, the freed-up memory can instead be utilized to double the context from 1024 to 2048 tokens.
In this paper, we introduce a direct transfer datapath designed to efficiently load the selected embedding vector from external flash into the LLM accelerator with acceptable latency.

\subsection{Grouped-query Attention in LLM}


Recent research has demonstrated that GQA offers a favorable trade-off between the performance of Multi-Head Attention (MHA) and the reduced $kv$ cache storage requirements of Multi-Query Attention (MQA) \cite{ainslie2023gqa}. In GQA, attention heads are partitioned into several groups, with all heads within a group sharing the same $kv$ cache while maintaining distinct query vectors. This design introduces data reuse opportunities compared to the MHA, thereby necessitating a distinct dataflow design to efficiently exploit this reuse. In this paper, we introduce a GQA dataflow that can efficiently handle such reuse pattern with acceptable resource consumption.

\subsection{Classic Embedded FPGA System}

\begin{figure}
  \centering
  \includegraphics[width=1.0\linewidth]{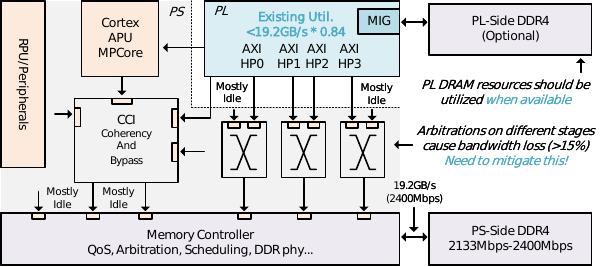}
  \caption{A classic Zynq UltraScale+ SoC with key component connections; PL memory access to PS DRAM via multiple AXI ports exists bandwidth loss.}
  \label{fig:zynq}
  \end{figure}

The Zynq UltraScale+ (ZU+) FPGA, as shown in Fig.\ref{fig:zynq}, represents a typical embedded FPGA with a heterogeneous System-on-Chip (SoC) that integrates CPUs on the processing system (PS) side with programmable logic (PL) for custom hardware acceleration. DRAM resources are typically connected to the PS and primarily serve the CPU, while memory access from the PL is provided via several 128-bit AXI ports. KV260 employs such PS-only DRAM configuration. Also, additional DRAM resources can be directly connected to the PL via high-speed I/O interfaces, as seen in platforms such as the ZCU104. In such cases, the PL can access dedicated memory using the Memory Interface Generator (MIG) IP core, which exposes a 512-bit AXI port to the user logic.

In this work, we focus on maximizing DRAM utilization in ZU+ systems for memory-bound LLM inference. ZU+ supports 64-bit DRAM speeds up to 2400 Mbps. The KV260 adopts this setup, offering a peak bandwidth of 19.2GB/s (2400Mbps$\times$ 64/8 = 19.2GB/s). 
Four 128-bit AXI ports running at 300MHz are needed to fully expose the bandwidth to the PL-side. However, we observe a performance inefficiency—a $1+1+1+1<4$ problem—where arbitration among the 4 AXI ports at the memory controller results in bandwidth loss.
This phenomenon has been recognized in prior studies.
Existing bandwidth stress-testing research \cite{manev2019unexpected, brilli2022understanding} shows that only 75-82\% of the theoretical bandwidth can be achieved due to resource contention among AXI ports sharing the same memory space.
Li et al.\cite{li2025pushing} reported a similar observation of 84\% bandwidth utilization in their LLM accelerator implementation.
In this work, we aim to mitigate this bottleneck through a novel memory access strategy.

Furthermore, in embedded FPGA platforms equipped with dedicated PL-side memory, it is advantageous to leverage this memory for LLM inference to improve throughput. To this end, we introduce a workload partitioning strategy that exploits the heterogeneous memory interfaces of the PS and PL, thereby achieving higher inference performance.

\section{Architecture Overview}

\begin{figure}
  \centering
  \includegraphics[width=1.0\linewidth]{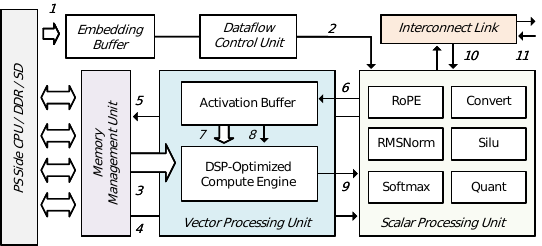}
  \caption{Architecture overview of Hummingbird. Connections between the PS, MMU, VPU and SPU are labeled (1-11) and explained in the main text.}
  \label{fig:arch}
  \end{figure}

  \begin{figure}
    \centering
    \includegraphics[width=1.0\linewidth]{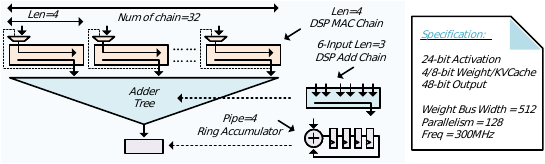}
    \caption{The proposed DSP-optimized GEMV compute engine.
    }
    \label{fig:gemv}
    \end{figure}

    \begin{figure*}
      \centering
      \includegraphics[width=1.0\linewidth]{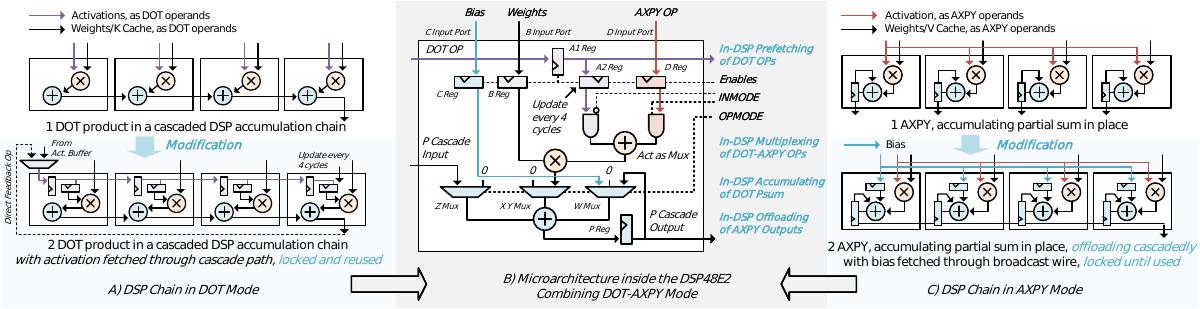}
      \caption{The proposed DSP MAC chain design supporting the (A) DOT and (C) AXPY mode, with internal microarchitecture of the DSP48E2 shown in (B).}
      \label{fig:dotaxpychain}
      \end{figure*}

An overview of the Hummingbird hardware architecture is illustrated in Fig.\ref{fig:arch}, highlighting several key components. The Memory Management Unit (MMU) is responsible for coordinating both on-chip and off-chip memory access, including buffering of the key-value ($kv$) cache. The Vector Processing Unit (VPU) executes dense General Matrix-Vector (GEMV) operations, which are central to LLM inference. The dense GEMV computation adopts 24-bit fixed point integer (INT24), and the results are converted to 16-bit floating point (FP16) for processing in the Scalar Processing Unit (SPU). SPU handles element-wise operations such as Rotary Position Embedding (RoPE)\cite{su2024roformer}, softmax, Layer Normalization\cite{xu2019understanding}, SiLU activation\cite{elfwing2018sigmoid}, and quantization. The Dataflow Control Unit (DCU) orchestrates system-wide data movement and computational flow across modules. The embedding buffer functions as a FIFO, receiving embedding vectors from the PS. Finally, the interconnect link allows communication among cores when tensor parallelism is enabled.

The labeled connections in Fig.\ref{fig:arch} are further described as follows:
(1) Embedding vector transfer from external flash memory to the embedding buffer.
(2) Datapath from the embedding buffer to the SPU for LayerNorm computation and residual buffering.
(3) Transfer of model weights and key-value ($kv$) cache data to the VPU for GEMV computation.
(4) Transfer of LayerNorm parameters and weight scales to the SPU for dequantization and normalization operations.
(5) Offloading of computed $kv$ cache back to external memory for future reuse during inference.
(6) SPU output transfer to the activation buffer, storing intermediate activation results as operands.
(7) Sending vector operands from the activation buffer to the VPU for DOT-product GEMV execution.
(8) Sending scalar operands from the activation buffer to the VPU for AXPY-style GEMV operations.
(9) Output from the VPU to the SPU for post-processing, including auxiliary computations such as non-linear activation functions.
(10) All-reduce synchronization across multiple cores to support tensor parallelism and maintain data consistency.
(11) Inter-core communication datapath enabling collective operations and coordination among parallel compute units.
The proposed architectural optimizations are elaborated in the next section.


\section{Optimizations}

\subsection{Smaller: DSP-Optimized Compute Engine}

The compute engine within the VPU is responsible for dense GEMV, the computational core of LLM decoding.
Different from prior FP16 designs\cite{li2025pushing}, we adopt a INT24 implementation to enable more fine-grained DSP optimizations\cite{li2023firefly}.

\textit{In our design, we strike a balance between the two classic GEMV hardware design}: the DSP-efficient yet flip-flop-intensive (needed for systolic data alignment) full DSP \textbf{\textit{chain}} architecture, and the flip-flop-efficient but DSP-inefficient MAC \textbf{\textit{tree}} architecture, which employs separate DSPs for multiplication and addition.
To achieve this, we propose a segmented DSP chain hybrid architecture. In this design, the full computation chain is partitioned into smaller DSP sub-chains, each performing partial accumulation. The outputs of these sub-chains are subsequently aggregated using an adder tree with a reduced number of inputs. This adder tree is also composed of multi-input DSP sub-chains, as shown in Fig.\ref{fig:gemv}.

\textit{While the hybrid implementation is straightforward, our design targets several additional optimizations:}
(1) Reduction of activation data bandwidth to the compute engine by exploiting data reuse in GEMV.
(2) Support for flexible GEMV computation modes, including both DOT and AXPY operations—the latter of which avoids $v$ transposition in attention mechanisms.
(3) Preservation of resource efficiency, ensuring that the above optimizations are achieved with minimum resource overhead.

\begin{figure}
  \centering
  \includegraphics[width=1.0\linewidth]{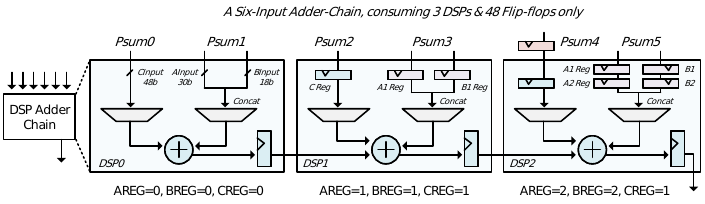}
  \caption{A six-input adder chain as the building block of the reduction tree.}
  \label{fig:addertree}
  \end{figure}

  \begin{table}[t]
    \centering
    \caption{Comparison of Resource Utilization of the Compute Engine}
    \begin{tabular}{lccccc}
    \toprule
    \textbf{Architecture} & \textbf{LUT} & \textbf{FFs} & \textbf{DSP} & \textbf{Act. Reuse} & \textbf{AXPY}\\
    \midrule
    FP16 MAC Tree              &    31872          &    44809  &  256         & No      & No   \\
    INT24 MAC Tree              &    0          &    9936      &  256         & No      & No   \\
    INT24 MAC Chain             &    0          &    59535     &  128         & No      & No   \\
    Hybrid                &    0          &    8784      &  160         & No      & No   \\
    Hybrid+ (w/o Opt.) &      6570       &    11856          &  160       & Yes        & Yes   \\
    Hybrid+ (w/ Opt.) &      1962        &   \textbf{4355}$\downarrow$           &  \textbf{148}$\downarrow$       & \textbf{Yes}        & \textbf{Yes}   \\
    \bottomrule
    \end{tabular}
    \label{tab:mac_utilization}
    \end{table}

\subsubsection{\textbf{Flexible DSP MAC Chain}} A straightforward implementation of the DSP MAC chain for DOT-product GEMV is illustrated in Fig.\ref{fig:dotaxpychain}A1. In this configuration, each DSP receives a distinct activation-weight pair, and the resulting products are accumulated sequentially along the chain. In GEMV scenario with activation reuse opportunities, the activation inputs can be reused across multiple cycles, thereby reducing the required activation data bandwidth. This enables the use of a shared activation data path feeding all DSPs, rather than individual activation inputs per DSP. Modern DSP blocks inherently support input cascading, facilitating activation prefetching along the chain, as shown in Fig.\ref{fig:dotaxpychain}A2. In the example of a chain with $Len=4$, the activation value is refreshed every 4 cycles.

Alternatively, GEMV can be implemented using an AXPY-style computation, where accumulation is performed in-place rather than being propagated along the cascaded chain, as depicted in Fig.\ref{fig:dotaxpychain}C1. In this approach, each DSP accumulates its partial result locally, which necessitates collecting outputs from all individual DSPs. This results in complex routing and increased resources consumption. To mitigate this, we can utilize the DSP output cascade path for efficient data offloading once the accumulation is complete, as shown in Fig.\ref{fig:dotaxpychain}C2.

\begin{figure*}
  \centering
  \includegraphics[width=1.0\linewidth]{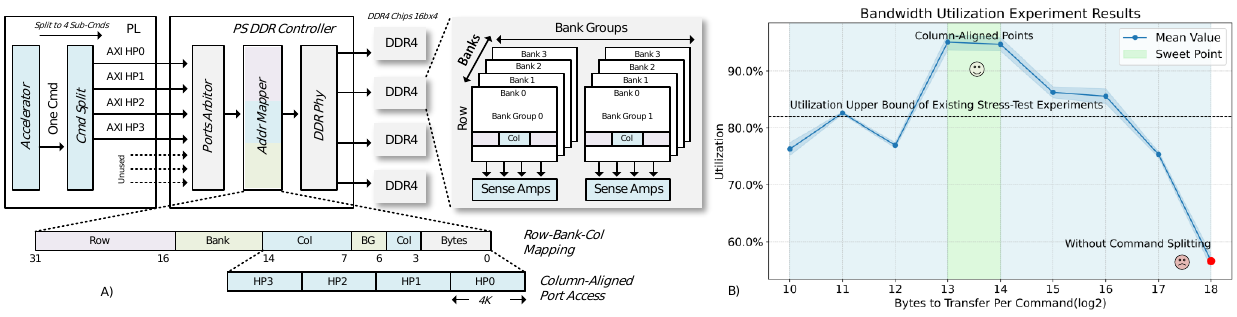}
  \caption{(A) PL-PS DDR access through multi AXI ports in Zynq platform, the hierarchical organization of the DDR4 chip, and the proposed column-aligned memory access strategy. (B) Bandwidth utilization of different BTT per transfer with the column-aligned sweet point region highlighted. }
  \label{fig:access}
  \end{figure*}

We demonstrate that both DOT-product and AXPY modes of GEMV can be supported and dynamically switched through flexible control of the Xilinx UltraScale DSP48E2 primitives, as shown in Fig.\ref{fig:dotaxpychain}B. This flexibility enables the application of several in-DSP optimization techniques\cite{li2024revealing}:

\textbf{In-DSP prefetching.} The DSP48E2 primitive provides individual clock-enable signals for its input registers. This allows manual control over when each internal register should be updated. By utilizing the dual-register configuration along the $A$ input cascade path, the $A1$ registers can serve as a data prefetching path, while the $A2$ registers can lock activations for reuse. This design eliminates the need for external routing or flip-flops, forming a fully in-DSP data path for activations.

\textbf{In-DSP Multiplexing.} The DSP48E2 includes a pre-adder that computes the sum of the $A$ and $D$ input data, both of which are gated by individual \texttt{AND} gates. These gates, together with the pre-adder, can function as a multiplexer, selecting between DOT operands (from the $A$ input) and AXPY operands (from the $D$ input). The AND gates are dynamically controlled using the DSP48E2's \texttt{INMODE} configuration. This approach requires no additional fabric resources, allowing the entire dynamical switching to occur within the DSP block.

\textbf{In-DSP Accumulation.} Cascaded DSP adder chains have been widely used in prior optimizations~\cite{samajdar2019scaling}\cite{wu2017high}. We leverage this structure to accumulate partial sums efficiently along the DSP chain during DOT computation.

\textbf{In-DSP Offloading.} While AXPY accumulation is performed in-place, the final output can be offloaded reusing the same the output cascade path. During offloading, the $Z$ multiplexer is activated only after accumulation is complete, while the $X$, $Y$, and $W$ multiplexers are disabled to prevent interference from incoming data. The selection signals for these multiplexers ($W$ $X$ $Y$ $Z$) are controlled through the \texttt{OPMODE} port of the DSP48E2.

These optimizations ensures flexible support for both DOT and AXPY GEMV, while simultaneously minimizing resource usage. \textit{It's worth noting that these fine-grained techniques cannot be automatically inferred by the toolchain and must be applied through manual instantiation of primitives.}

\subsubsection{\textbf{DSP-Efficient Reduction Tree}}
Each DSP48E2 can function as a standalone 48-bit adder, with one operand from the $C$ input port, and the other formed by concatenating the 30-bit $A$ input and the 18-bit $B$ input within the DSP.
Again, we strike a balance between the tree and chain implementation and construct a $Len=3$ adder chain which has minimum number of required flip-flops shown in Fig.\ref{fig:addertree}, as the basic building block of the reduction tree.
We exploit the configurable pipeline depths of the DSP48E2 input paths to reduce fabric flip-flops. The $A$ and $B$ input pipelines can be configured to depths of 0, 1, or 2 stages, while the $C$ input pipeline supports depths of 0 or 1. By tailoring these pipeline configurations to match the systolic latency, fabric register usage can be minimized. As shown in Fig.\ref{fig:addertree}, this approach requires only a single set of 48 fabric flip-flops. The resulting 6-input pipelined adder chain serves as the fundamental unit of the adder tree, thus achieve less DSP usage compared to adder tree implemented using 2-input individual DSP adders.

We conduct ablation experiments to evaluate the area efficiency of the proposed DSP-optimized compute engine shown in Table.\ref{tab:mac_utilization}. Results show that the hybrid design balances tree and chain implementations, while our enhanced version—with activation prefetching and AXPY support—achieves significant resource savings over the unoptimized baseline, with acceptable LUT overhead compared to the basic hybrid design.

\begin{table}[t]
  \centering
  \caption{Comparison of Model Weight Transfer Latency on KV260}
  \begin{tabular}{lcc}
  \toprule
  \textbf{Metric} & \textbf{LLaMA2-7B} & \textbf{LLaMA3-8B}\\
  \midrule
  Model Size(Per Inference) &    3249MB     &    3690MB      \\
  Latency (w/o Opt.)        &    201ms (5 pass/s)      &    228ms (4.3 pass/s)       \\
  Latency (w Opt.)          &    178ms$\downarrow$(5.6 pass/s$\uparrow$)      &    202ms$\downarrow$(5 pass/s$\uparrow$)       \\
  \bottomrule
  \end{tabular}
  \label{tab:model_transfer}
  \end{table}

\subsection{Faster: Column-Aligned Memory Access}

We propose a novel memory access strategy to address the $1+1+1+1<4$ problem, wherein the combined bandwidth of four AXI ports reaches only 84\% of the theoretical bandwidth\cite{li2025pushing} due to arbitration loss.
As arbitration is unavoidable in a fixed SoC, optimizing the access pattern becomes essential for efficiency. Specifically, when accessing a large weight matrix through multiple AXI ports, it is critical to determine the optimal workload granularity per command assigned to each port, balancing burst transfer efficiency with access locality.

To address this, we examine the internal organization of the DDR4\cite{sohn20121} chip, as shown in Fig.\ref{fig:access}A. Typically, DRAM is structured hierarchically into bank groups, banks, rows, and columns. Each bank comprises a 2D row-by-column array, and the physical address of the memory is correspondingly partitioned into bank group, bank, row, and column components.
When an access command is issued, the memory controller first decodes the bank group and bank to identify the target array. The row address then activates a word line, and the column address selects the desired portion of that word line, which is read out via sense amplifiers. If a subsequent access targets the same row (which remains open), the read can proceed immediately with minimal latency. However, if the new access targets a different row, the current row must first be closed with a precharge command before the new row is activated—introducing additional latency. This row switching overhead is a key contributor to bandwidth loss in non-sequential access patterns.
Careful organization of physical addresses can mitigate this by enabling bank multiplexing and interleaving. These techniques allow overlapping of precharge and activate operations across different banks, effectively concealing latency, as long as access locality is maintained.

Given the DRAM mechanisms, it is evident that bandwidth loss primarily arises from unpredictable command arbitration which disrupts the original sequential and locality-preserving access pattern and leads to unnecessary row and bank switching. 
Based on this observation, we propose the following heuristic: during each transaction, the data accessed by all 4 ports should reside within the same row and bank, with each port accessing a distinct column segment. Under this condition, the access order of the ports becomes irrelevant, as all ports operate on the same open row, thus avoiding row or bank switching.
To satisfy this constraint, the base address of each transaction (before being split across 4 ports) must be aligned with the column boundary, and the bytes-to-transfer (BTT) per transaction must not exceed the column size. 

We conduct experiments to determine the optimal BTT per transaction on both KV260 and ZCU104 as shown in Fig.\ref{fig:access}B.
We test the bandwidth utilization of transferring a 256KB matrix split into multiple transactions with the BTT per transfer ranging from $2^{10}$ to $2^{18}$ bytes. Fig.\ref{fig:access}B shows that peak bandwidth could not be reached when BTT is too small (inadequate burst lengths) or too large (causing row/bank switching). Optimal BTT is achieved when transaction size matched the column size or the grouped column size, specifically $2^{13}$ or $2^{14}$ on both KV260 and ZCU104.
We adopt this column-aligned access strategy on Zynq FPGAs, achieving up to 95\% bandwidth utilization—10\% higher than previously reported in bandwidth stress-test studies on Zynq\cite{manev2019unexpected, brilli2022understanding} and the SOTA embedded LLM accelerator\cite{li2025pushing}.
This bandwidth utilization enhancement directly contributes to a faster weight transfer speed shown in Table.\ref{tab:model_transfer}.
It is worth noting that additional 1-2\% utilization drop may occur during actual inference.


\subsection{Faster: Tensor Parallelism for Bandwidth Scale-Up}

Megatron LM\cite{shoeybi2019megatron} introduces a tensor parallelism strategy designed to distribute LLM inference workloads across multiple GPUs. By leveraging the structural features of transformers\cite{vaswani2017attention}, Megatron LM minimizes synchronization points between compute units while partitioning the workload.
In the attention layer, computations are distributed across multiple heads, allowing each compute unit to process attention independently within its assigned head. In the MLP layer, the upward projection is partitioned along the output dimension, while the downward projection is split along the input dimension. All-reduce synchronization occurs only before the normalization process and can be conducted concurrently with the attention output projection and the MLP downward projection.

On the ZCU104 platform, which provides additional DDR memory on the PL side, a dual-core design can be employed to fully exploit the available bandwidth and memory capacity for LLM decoding acceleration. Each core accesses a distinct memory region and independently executes a partitioned workload.
Megatron tensor parallelism is inherently scalable. A four-core system can be effectively deployed on Alveo U250, each equipped with four independent DDR4 memory banks. The all-reduce process is effectively hidden behind the GEMV computation, thus minimizing synchronization overhead, enabling linear performance scaling.

\subsection{Embedding Offloading and Direct Transfer Datapath}

\begin{figure}
  \centering
  \includegraphics[width=1.0\linewidth]{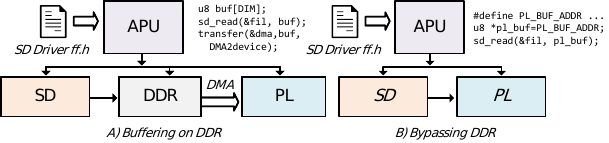}
  \caption{Bypassing DDR for embedding vector transfer}
  \label{fig:sdtransfer}
  \end{figure}

  \begin{table}
    \centering
    \caption{Averaged Embedding Vector Loading Latency}
    \begin{tabular}{lccc}
    \toprule
    \textbf{Metric} & \textbf{w/o Opt.} & \textbf{+FastSeek, Buffering} & \textbf{+FastSeek, Bypassing}  \\
    \midrule
    Latency           & 152ms &  2.8ms &  1.5ms$\downarrow$ \\
    \bottomrule
    \end{tabular}
    \label{tab:sdload}
    \end{table}

We evaluate the latency of loading embedding vectors from an SD card and find that enabling the fast seek feature significantly reduces load time. This feature builds a cluster link map during an initial scan of the file's cluster chain, allowing subsequent seeks to jump directly to the target cluster, which is especially useful for locating embedding vectors.
To further cut latency, we introduce a DDR-bypassing method. Unlike the two-stage approach—where data is first loaded into PS DDR and then transferred via DMA to the PL—our method allocates a dedicated address region on the PL, enabling direct SD-to-PL transfers and eliminating intermediate buffering and DMA overhead, as shown in Fig.\ref{fig:sdtransfer}. The fast seek feature and the bypassing technique improves loading efficiency. Latency comparisons are shown in Table.\ref{tab:sdload}.

\subsection{GQA Dataflow Design}

\begin{figure}[]
  \centering
  \includegraphics[width=1.0\linewidth]{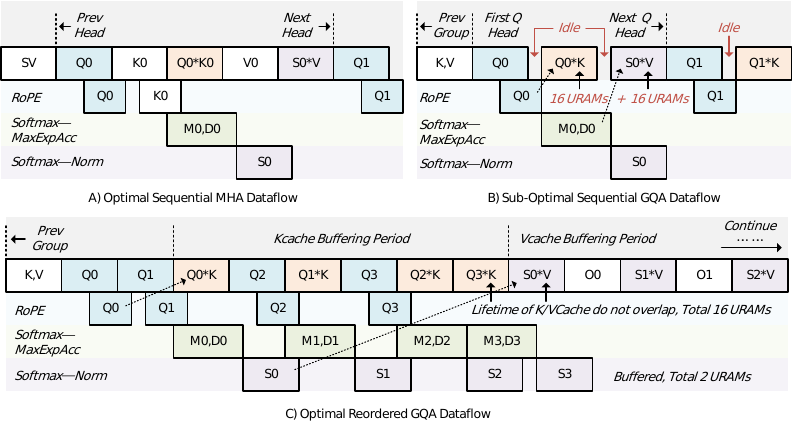}
  \caption{Illustration of (A) MHA (B) sub-optimal GQA and (C) proposed GQA dataflow. The dash line indicates the data dependency.}
  \label{fig:gqadataflow}
  \end{figure}

Prior work\cite{li2025pushing} introduced an MHA dataflow that effectively hides the latency of RoPE and softmax behind the dense GEMV computation, as shown in Fig.\ref{fig:gqadataflow}A.
However, a naïve extension of this dataflow to GQA fails to fully hide the these latency and results in idle periods shown in Fig.\ref{fig:gqadataflow}B.
To address this, we propose an optimized GQA dataflow shown in Fig.\ref{fig:gqadataflow}C, with a group size of 4 as an example.

Two major inefficiencies in the original GQA dataflow are addressed in our design.  
The first inefficiency arises from the RoPE operation. Although RoPE computation can begin once half of the $q_i$ vector is generated, it requires the full $q_i$ vector to perform the paired rotation\cite{li2025efficient}. As a result, the subsequent computation of $q^R_i \cdot [K_{\text{Cache}}, k^R]$ must wait until $q^R_i$ is fully available, introducing idle cycles (78 tail cycles needed by RoPE in our implementation). To mitigate this, we reorder the pipeline by placing two query computations ahead. This ensures that $q^R_0$ is ready in time for the corresponding RoPE and GEMV operations, effectively eliminating the idle period.


The second inefficiency is related to memory usage for $kv$ cache. In GQA, $k$ and $v$ are reused across multiple $q$ heads, therefore on-chip buffering is preferred to reduce memory access. For LLaMA3-8B, buffering 4096 8-bit $k$ or $v$ of dimension 128 requires 16 URAMs, which is acceptable for KV260 and ZCU104 with 64 and 96 URAMs, respectively.
However, a naïve GQA dataflow would require simultaneous storage of both $k$ and $v$ caches, doubling the memory requirement to 32 URAMs—posing a significant overhead.
To address this, we restructure the dataflow to first compute all $q^R_i K$ results and store the softmax outputs, which require far less memory (requiring only 2 extra URAMs). These stored softmax results $s_i$  are then reused for the subsequent $s_i V$ computation. This approach not only reduces URAM usage but also effectively hides the softmax latency.
It is worth noting that we adopt an online softmax\cite{milakov2018online} FPGA implementation\cite{he2024fpga}, which fuses the maximum value search and exponential accumulation passes into a single pass. Such approach introduces a fixed number of tail cycles during the first pass (47 cycles in our implementation), and it is efficiently hidden by our proposed GQA dataflow. After each $s_i  V$ computation, instead of globally collecting AXPY results from all DSP chains, AXPY results are locally routed to the activation loading ports of their respective DSP chains, directly acting as the activation input of the subsequent output projection computation for the current head. This avoids global collection, reducing fan-out pressure and eliminating serialization latency.

\section{Experiment}
\subsection{Hardware Setup}

\begin{figure}[]
  \centering
  \includegraphics[width=1.0\linewidth]{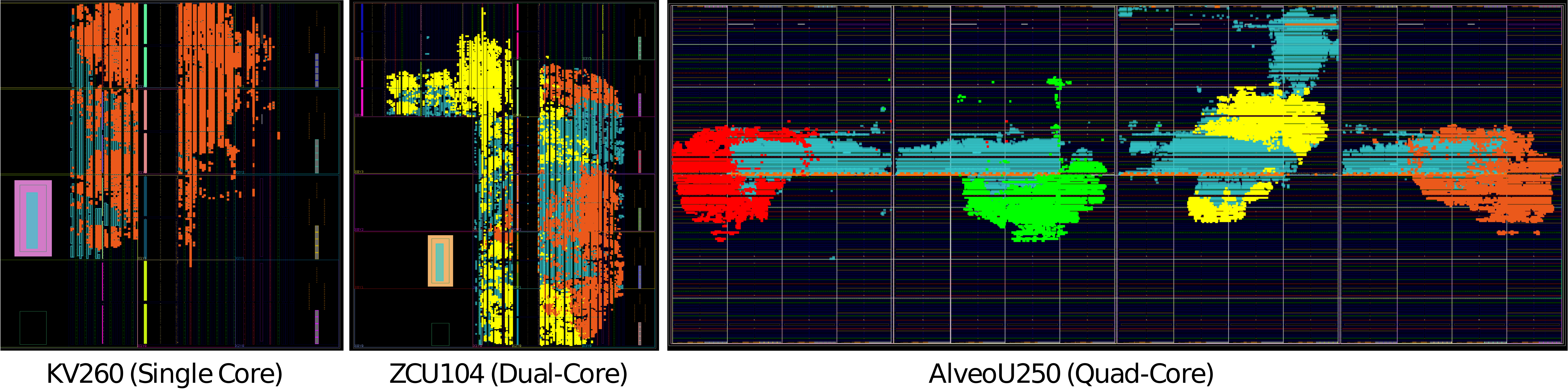}
  \caption{Hummingbird's layout with different colors indicating different cores.}
  \label{fig:layout}
  \end{figure}

\subsubsection{Implementation}
Hummingbird is implemented using SpinalHDL\cite{spinalhdlcite}, with each module verified through Verilator\cite{snyder2004verilator} and implemented in Vivado. Resource and power consumption metrics are derived from Vivado reports.

\subsubsection{Platforms}
As shown in Fig.\ref{fig:layout}, Hummingbird is deployed on:
(1) KV260, with 4GB of 2400Mbps 64-bit DDR4 at PS, providing 19.2GB/s.
(2) ZCU104, with 2GB of 2133Mbps 64-bit DDR4 at PS side and an plug-in DIMM of 2133Mbps 64-bit DDR4 at PL, yielding a total bandwidth of 34GB/s.
(3) To evaluate the scalability of our design, we also implement Hummingbird on Alveo U250, with four 2400Mbps 64-bit DDR4 across four SLRs, providing 76.8GB/s in total.

\subsubsection{LLM}
Hummingbird supports LLaMA3-8B\cite{grattafiori2024llama3herdmodels}, which is symmetrically quantized to 4 bits using GPTQ\cite{frantar2022gptq}, while the $kv$ cache is quantized to 8 bits using simple linear quantization. No extra compression techniques are applied.
While LLaMA3-8B supports up to 8K context, this exceeds the memory limits of KV260 and ZCU104 even with embedding offloaded. Notably, our 4K context support is $4\times$ larger than the 1K handled by the SOTA embedded LLM accelerator\cite{li2025pushing}.

\subsection{Resource Breakdown}

A detailed breakdown of resource consumption for the standalone Hummingbird IP is presented in Table.\ref{tab:breakdown}. The VPU accounts for the majority of compute resource usage, while the MMU dominates memory resource consumption.

\begin{table}[H]
  \centering
  \caption{Resource Consumption Breakdown of the Hummingbird IP}
  \begin{tabular}{lccccc}
  \toprule
    & \textbf{LUT} & \textbf{FF} & \textbf{BRAM} & \textbf{URAM} & \textbf{DSP} \\
  \midrule
  VPU           & 2859   &  5173   &  12 & 0  & 150  \\
  SPU           & 5790   &  6811   &  11 & 0  & 29   \\
  MMU           & 7670   &  9232   &  34 & 18 & 0  \\
  Total         & 18355  &  25422  &  59 & 18 & 179  \\
  \bottomrule
  \end{tabular}
  \label{tab:breakdown}
  \end{table}

\subsection{Comparison Results}

\begin{table*}[t]
  \centering
  \setlength{\tabcolsep}{5pt}
  \caption{Performance and Metrics Comparison of LLM Accelerators on Different FPGA platforms}
  \begin{tabular}{lccccccccccc}
  \toprule
  \multicolumn{1}{c}{} & \multicolumn{3}{c}{\textbf{HBM-Based Cloud FPGA}} & \multicolumn{4}{c}{\textbf{DDR-Based Cloud FPGA}} & \multicolumn{4}{c}{\textbf{DDR-Based Embedded FPGA}} \\
  \cmidrule(lr){2-4} \cmidrule(lr){5-8} \cmidrule(lr){9-12}
  & FlightLLM & Chen\cite{chen2024understanding} & LoopLynx & \multicolumn{2}{c}{ChatOPU} & LightMamba & \textbf{\textit{Hummingbird}} & Li\cite{li2025pushing} & LlamaF & \multicolumn{2}{c}{\textbf{\textit{Hummingbird}}} \\
  \midrule
  {Device}      & U280      & U280       & U50       & \multicolumn{2}{c}{U200}     & VCK190     & U250        & KV260    & ZCU102    & ZCU104     & KV260             \\
  {GB/s}        & 460       & 460        & 201       & 19.2   & 76.8                & 12/76.8    & 76.8        & 19.2     & 19.2      & 34.1       & 19.2              \\
  \midrule
  {LUT}         & 574K      & 569K       & 312K      & 755K   & $4\times$           & 107K       & 168K        & 78K      & 164K      & 66K$\downarrow$        & 26K$\downarrow$               \\
  {FF}          & 943K      & 653K       & 478K      & 998K   & $4\times$           & 130K       & 218K        & 105K     & 171K      & 92K$\downarrow$        & 35K$\downarrow$               \\
  {DSP}         & 6345      & 1780       & 1132      & 1091   & $4\times$           & 228        & 771         & 291      & 528       & 362        & 179$\downarrow$               \\
  {B/URAM}      & 1252/792  & 389/111    & 924.5/4   & 417/540& $4\times$           & 912/61     & 361/76      & 37/10    & 223/-     & 133/40     & 59/18$\downarrow$             \\
  {MHz}         & 225       & 245        & 285       & 300    & 300                 & 400        & 300         & 300      & 205       & 266        & 300               \\
  {W}           & 45        & 29.5       & 39.4      & -      & -                   & 3.2        & 14.61       & 6.57     & 5.08      & 7.09       & 3.81$\downarrow$              \\
  \midrule
  {Tasks}       & llama2-7b & gpt2-345m  & gpt2-345m & \multicolumn{2}{c}{opt-350m} & mamba2-2.7b& llama3-8b$\uparrow $   & llama2-7b& llama-1.1b& \multicolumn{2}{c}{llama3-8b$\uparrow $}  \\
  {Opt.}        & SparseW8  & W8         & W8        & \multicolumn{2}{c}{SparseW16}& W4         & W4          & W4       & W8        & W4         & W4                \\
  {Token/s}     & 55        & 204        & 260       & 43.2   & 166.2               & 7.21       & 19.4        & 4.9      & 1.48      & 8.6$\uparrow $        & 4.8               \\
  {Norm.Perf.}  & 55        & 19.6       & 25        & 4.2    & 16.2                & 2.78       & 22.17$\uparrow $       & 4.9      & 0.47      & 9.83$\uparrow $       & 5.48$\uparrow $              \\
  {BW. Eff.}    & 65\%      & 23\%       & 59\%      & 72\%   & 66\%                & 75\%/12\%  & 91\%$\uparrow $        & 84\%     & 8\%       & 93\%$\uparrow $       & 94\%$\uparrow $                  \\
  {PW. Eff.}    & 1.22      & 0.66       & 0.63      & -      & -                   & 0.87       & 1.51$\uparrow $        & 0.74     & 0.09      & 1.39$\uparrow $       & 1.44$\uparrow $              \\
  \bottomrule
  \end{tabular}
  \label{tab:fpga_comparison}
\end{table*}

\begin{table}
  \centering
  \caption{Comparison with Jetson Orin series GPUs}
  \begin{tabular}{lcccccc}
  \toprule
    & \textbf{GB/s} & \textbf{Arch} & \textbf{token/s} & \textbf{BW.Eff.} & \textbf{PW.Eff.} \\
  \midrule
  OrinNano   & 68    &  NanoLLM      &  15\cite{jetsonbenchmark} & 79\% & 1.0 \\
  OrinAGX    & 204.8 &  NanoLLM      &  40\cite{jetsonbenchmark} & 71\% & 0.66\\
  KV260            & 19.2  &  Hummingbird  &  4.8  & 94\%$\uparrow $   & 1.44$\uparrow $  \\
  ZCU104           & 34.1  &  Hummingbird  &  8.6  & 93\%$\uparrow $   & 1.39$\uparrow $  \\
  \bottomrule
  \end{tabular}
  \label{tab:gpucompare}
  \end{table}

  \begin{table}
    \centering
    \caption{Hummingbird on Spartan UltraScale+ FPGA}
    \begin{tabular}{lccccc}
    \toprule
      & \textbf{LUT} & \textbf{FF} & \textbf{BRAM} & \textbf{URAM} & \textbf{DSP} \\
    \midrule
    SU150P           & 63K  &  126K  &  168 & 16 & 384  \\
    SU200P           & 100K &  200K  &  192 & 64 & 384  \\
    Hummingbird(1-Core)      & 18K  &  25K   &  59(+16)  & 18(-2) & 179  \\
    Hummingbird(2-Core)      & 34K  &  47K   &  108 & 40 & 359  \\
    \bottomrule
    \end{tabular}
    \label{tab:Spartan}
    \end{table}

Table.\ref{tab:fpga_comparison} shows the comparison between Hummingbird and existing LLM accelerators on DDR-based embedded FPGAs (Li et al.\cite{li2025pushing}, LlamaF\cite{xu2024llamaf}), DDR-based cloud FPGAs (ChatOPU\cite{chatgptcite}, LightMamba\cite{wei2025lightmamba}) and HBM-based cloud FPGAs (FlightLLM\cite{zeng2024flightllm}, Chen et al.\cite{chen2024understanding}, LoopLynx\cite{zheng2025looplynx}).
The table summarizes key system and performance characteristics, including device type, bandwidth, resource and power consumption, operating frequency, and task-specific performance metrics.
All token/s metrics reflect single-batch decoding speed under a short-context setting, where the $kv$ cache size is negligible compared to model weights.
We report the decoding speed under a prefill:decode = 32:32 setup.
Normalized performance (\textit{Norm. Perf.}) is the normalized token/s throughput with model size normalized to 7B and compression strategy normalized to dense 4-bit format.
Bandwidth efficiency (\textit{BW. Eff.}), as defined in \cite{agarwal2023llm}, refers to the ratio between the observed token/s and the number of times the model weights could theoretically be transferred across the available memory bandwidth per second. This metric reflects the efficiency of the accelerator since LLMs are typically memory-bound.
Power efficiency (\textit{PW. Eff.}) is defined as the \textit{Norm. Perf.} per watt.

As shown in Table.\ref{tab:fpga_comparison}, Hummingbird is the only accelerator capable of deploying LLMs larger than 7B on a single FPGA. Across all three evaluated FPGAs, Hummingbird demonstrates consistently low resource consumption compared to existing accelerators across various categories, while achieving superior bandwidth and power efficiency.
On the KV260, Hummingbird achieves the lowest resource utilization and the highest bandwidth and power efficiency among all accelerator categories. The decoding speed for LLaMA3-8B on the KV260 is slightly lower than that of Li et al.\cite{li2025pushing} (4.8 vs. 4.9 tokens/s), primarily because Hummingbird targets a 10\% larger model (8B vs. 7B).
On the ZCU104, Hummingbird achieves the fastest decoding speed among all DDR-based embedded FPGA accelerators. On the Alveo U250, it achieves the highest normalized performance among DDR-based cloud FPGA accelerators.
Although we cannot match the normalized performance of HBM-based accelerators—due to significantly lower available memory bandwidth, we still outperforms all such designs in terms of bandwidth and power efficiency.

\subsection{Comparison with Jetson GPUs}

Table.\ref{tab:gpucompare} compares decoding performance and bandwidth efficiency for the 4-bit quantized LLaMA3-8B between Hummingbird on the KV260/ZCU104 and Jetson series embedded GPUs. Despite the KV260 and ZCU104 having significantly lower bandwidth, Hummingbird demonstrates superior bandwidth and power efficiency compared to Jetson series GPUs.

\subsection{Deployment on Emerging Cost-Optimized FPGAs}

\begin{figure}
  \centering
  \includegraphics[width=1.0\linewidth]{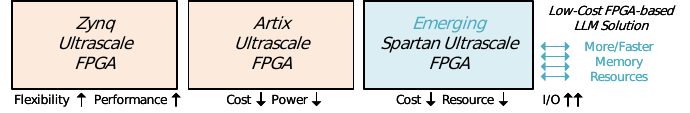}
  \caption{Cost-optimized Spartan UltraScale FPGA.}
  \label{fig:spartanfpga}
  \end{figure}

Spartan UltraScale+ is an emerging FPGA family that is cost-optimized with high-I/O ratio for edge applications (shown in Fig.\ref{fig:spartanfpga}), but with fewer resources than typical FPGA devices, it poses challenges for accelerator design (shown in Table.\ref{tab:Spartan}). \textit{\textbf{Hummingbird} is the \textbf{only} LLM accelerator small enough to fit within the resource constraints of SU150P and SU200P, making it the most cost-efficient LLM solution on embedded FPGAs.}
While official pricing is pending, SU+ FPGAs are expected to be significantly more affordable than existing FPGAs for low-cost LLM solution.

\section{Conclusion}

In this work, we propose \textbf{\textit{Hummingbird}}, a compact and high-performance LLM accelerator for embedded FPGAs, capable of supporting LLaMA3-8B.
It offers a cost-effective solution deployable on cost-optimized Spartan UltraScale+ FPGAs, demonstrating strong real-world applicability.

\section{Acknowledgement}
This work is supported by Grant No. E411230101 from Institute of Automation, Chinese Academy of Sciences.
We would also like to thank Renjie Wei and Jianing Zheng for providing statistics of the paper\cite{wei2025lightmamba}\cite{zheng2025looplynx} respectively.

\bibliographystyle{IEEEtran}
\bibliography{myIEEE,reference}

\end{document}